# A Leader-Follower Game Theoretic Approach to Arrest Cascading Failure in Smart Grid

Sohini Roy, Arunabha Sen
*School of Computing, Informatics and Decision System Engineering*
*Arizona State University*
Tempe-85281, Arizona, USA
Email: {sohini.roy, asen}@asu.edu

*Abstract*—The Smart Grid System (SGS) is a joint network comprising the power and the communication network. In this paper, the underlying intra-and-interdependencies between entities for a given SGS is captured using a dependency model called Modified Implicative Interdependency Model (MIIM) [1]. Given an integer K, the K-contingency list problem gives the list of K-most critical entities, failure of which maximizes the network damage at the current time. The problem being NP complete [2] and owing to the higher running time of the given Integer Linear Programming (ILP) based solution [3], a much faster heuristic solution to generate an event driven self-updating K-contingency list [4] is also given in this paper. Based on the contingency lists obtained from both the solutions, this paper proposes an adaptive entity hardening technique based on a leader-follower game theoretic approach that arrests the cascading failure of entities in the SGS after an initial failure of entities. The validation of the work is done by comparing the contingency lists using both types of solutions, obtained for different K values using the MIIM model on a smart grid of IEEE 14-Bus system with that obtained by simulating the smart grid using a co-simulation system formed by MATPOWER and Java Network Simulator (JNS). The K-contingency list obtained for a smart grid of IEEE 14-Bus system also indicate that the network damage predicted by both the ILP based solution and heuristic solution using MIIM are more realistic compared to that obtained using another dependency model called Implicative Interdependency Model (IIM) [2]. Advantage of using the MIIM based heuristic solution is also shown in this paper when larger SGS of IEEE 118-Bus is considered. Finally, it is shown how the adaptive hardening helps in improving the network performance.

NOMENCLATURE

*A. Power network entities*

All the power network entities are denoted as P type entities in this paper.
1) $P_a$: Bus with ID a.
2) $P_{a,b}$: Transmission Line between Bus a and Bus b.
3) $P_{BattX}$: Battery backup with ID X.

*B. Communication network entities*

All the communication network entities are denoted as C type entities in this paper. These C type entities are divided into 3 types–

1) **Type 1**: Substation entity, denoted as $(C_{1,X,Y,Z})$. The values of X,Y and Z depends on the following subdivisions.

    a) **Substation Server** $(C_{1,1,Y,Z})$: Y is the server ID and Z is the substation ID.

    b) **Substation Gateway** $(C_{1,2,Y,Z})$: Y is the gateway ID and Z is the substation ID.

    c) **LAN wire between server and gateway** $(C_{1,3,Y,Z})$: Y is the LAN wire ID and Z is the substation ID.

    d) **Optical fiber channel between SONET-Add-Drop Multiplexer (SADM) and substation gateway** $(C_{1,4,Y,Z})$: Y is the SADM ID and Z is the gateway ID.

    e) **Optical fiber channel between Optical-Add-Drop Multiplexer (OADM) and substation gateway** $(C_{1,5,Y,Z})$: Y is the OADM ID and Z is the gateway ID.

    f) **Communication channel between Remote Terminal Unit (RTU) and substation gateway** $(C_{1,6,Y,Z})$: Y is the RTU ID and Z is the substation ID.

    g) **Communication channel between Phasor Measurement Unit (PMU) and substation gateway** $(C_{1,7,Y,Z})$: Y is the PMU ID and Z is the substation ID.

2) **Type 2**: Synchronous Optical Networking Ring (SONET– Ring) entity, denoted as $(C_{2,X,Y,Z})$. The values of X,Y and Z depends on the following subdivisions.

    a) **SADM** $(C_{2,1,Y,0})$: Y is the SADM ID and Z=0 indicates that this type 2 entity is an SADM and not a connection.

    b) **Optical fiber channel between two SADMs** $(C_{2,2,Y,Z})$: Y is the first SADM ID and Z is the second SADM ID in the link.

3) **Type 3**: Dense Wavelength Division Multiplexing Ring (DWDM–Ring) entity, $(C_{3,X,Y,Z})$

    a) **OADM** $(C_{3,1,Y,0})$: Y is the OADM ID and Z=0 indicates that this type 3 entity is an OADM and not a connection.

    b) **Optical fiber channel between two OADMs** $(C_{3,2,Y,Z})$: Y is the first OADM ID and Z is the second OADM ID in the link.

*C. Entities connecting the P type entities to the C type entities*

Entities that cannot be identified as P type or C type entities are termed as connecting entities. These entities define the interdependencies between the two types of network entities.

1) $L_{1,i}$: Power supply line to a substation server where i is the ID of the line.
2) $L_{2,i}$ : Power supply line to a substation gateway where i is the ID of the line.
3) $L_{3,i}$ : Power supply line to an SADM and i is the ID of the line.
4) $L_{4,i}$ : Power supplying channel to an OADM where i is the ID of the line.
5) $L_{5,i}$ : Backup power supply line to a substation server from a substation battery where i is the ID of the battery.
6) $L_{6,i}$ : Back up power supply line to a substation gateway from a substation battery where i is the ID of the battery.
7) $U_i$ : Phasor Measurement Unit (PMU) with ID i.
8) $R_i$ : Remote Terminal Unit (RTU) with ID i.

## II. INTRODUCTION

The Smart Grid System (SGS) can be viewed as a two-layered network where one layer is composed of the power entities and the other layer is formed with communication entities. Yet, both the layers are connected to each other and the components of one layer depend highly on the components of the other layer for their operation. For example, the power system measurements of the smart grid obtained by its sensors must be transferred to the control center by the communication entities. Conversely, the communication entities themselves need power from the smart grid for their continued functionality. It should also be noted that the entities of each layer of the network also exhibit intra-dependencies among them. Therefore, if components of both the layers operate as required then only the SGS as a whole can function properly.

Now, due to this complex intra-and-interdependencies between the entities of two layers, if one or more entities in one layer of the SGS fail then other entities of both the layers will also fail as a result. Those newly failed entities will again initiate the failure of more entities. This is known as cascading failure [5] of entities. This cascading failure continues till a steady state is reached where this chain of failures breaks. Therefore, it is beyond any question that this cascading failure of entities can lead to a catastrophe where the whole SGS can fail. Thus, it is very essential to arrest the cascading failure as soon as it begins and protect the rest of the network from getting affected. In order to do that, the complex dependencies between the smart grid entities should be understood very well.

Identifying the need for clear understanding of the complex dependencies in a joint network, researchers made numerous efforts to come up with a model [6] that can vividly portray the smart grid system. Most of those models are too naïve to capture the complicated nature of interdependencies between the two kinds of networks in a smart grid. In [7] and [8] a high-level idea of the design of a joint network is given using a test system consisting of 14 buses. However, the ground level details of the Information and Communication Technology (ICT) network are missing in them. Moreover, the test systems presented in most of the papers differ a lot from the design of a real joint network. The Boolean logic-based implicative interdependency model (IIM) [2] overcame many of the afore-mentioned drawbacks. However, it also fails to accurately capture the communication network entities as it lacks knowledge of the communication network design. With the help of power utilities in the U.S. Southwest, the Modified Implicative Interdependency Model [1] presents a realistic design of the structure and operation of the power-and-communication network of a typical SGS. MIIM also considers different operational levels of the entities and models the complex dependencies between the two layers using multi-valued Boolean Logic based equations called Interdependency Relations (IDRs). In this paper an overview of the concepts of MIIM [1] are presented and those are used in order to arrest the cascading failures caused due to different attacks on the smart grid.

The SGS can face attacks from different types of attackers like the nature causing natural calamities, human causing physical or cyber-attacks and the smart grid itself can also act as an attacker to itself when cascading failures begin due to no external effects but the SGS entities themselves which fail to operate as desired. This paper proposes a novel approach that can defend any kind of attack to the SGS. This defense mechanism follows the Leader-Follower game theoretic concept [9]. Yet, in order to apply this technique, the Smart Grid Operators (SGOs) need a self-updating K-contingency list [4].

The set of entities, damage of which can result in the failure of maximum number of entities in the smart grid system are identified as the most critical set of entities. Upon this set of most vulnerable entities, the operability of the SGS is contingent and a list of such entities in the system is termed as the contingency list [4]. Usually, SGOs are provided with manuals that contain guidelines for handling different contingencies [10] in the system. Yet, in reality, when simulated contingencies do occur, the actual SGS may lie in a very different state than the simulations assumed. This results in either over-compensated or an under-compensated response. There comes the need for a measurement based self-updating contingency list which can provide real-time information to the operator about the current operational state of the entities. The current goal of the researchers is to find a suitable method to generate a Phasor Measurement Unit (PMU)-measurement based self-updating K-contingency list. In MIIM, each entity is associated with an operational state value of 0 indicating no-operation, 1 indicating reduced operation and 3 indicating full operation and these state values are updated each time a change in the operational level of an entity takes place. Such updating of operational values takes place on the basis of PMU data. Therefore, it can be stated that the state values of the entities in MIIM carry real-time information about the entities in the SGS. Ideally the self-updating contingency list for a given SGS can be identified just by solving the IDRs of the MIIM model each time a change takes place in the system. Efficient hardening [11] techniques followed for such critical entities can save the smart grid from a huge damage.

Yet, even after identifying all the vulnerable entities in the system, the smart grid operator can have a budget constraint of hardening only K entities of the network, where K can be any integer. In that case, it is important to identify the K-most critical entities in the system. The problem of identifying the K-most vulnerable entities in a SGS is already proved to be NP complete in [2]. Therefore, an Integer Linear Programming (ILP) based solution [3] for the problem is given in this paper using the MIIM IDRs. Now, every time an event of failure or recovery or reduced operation takes place in the SGS, the IDRs change, and the K-Contingency list keeps on changing. It becomes very challenging to update the MIIM IDRs and also generate the ILP based K-Contingency list within 33 ms (considering PMU data is obtained at 30 samples per second). Owing to the computation complexity of the problem, it is very difficult to come up with an accurate solution within that time span. Therefore, a much faster heuristic solution for generating a self-updating K-Contingency list within the given 33 ms is also given in this paper. Validation of the results from both the ILP and heuristic solutions is done by co-simulating the two layers of the smart grid network of IEEE 14-Bus system using MATPOWER and Java Network Simulator (JNS). A comparative study of the K-contingency list obtained using the MIIM IDRs is done with that obtained using IIM for a smart grid of IEEE 14-Bus system also. This paper also shows how the heuristic solution is beneficial in case of larger smart grids like that built using IEEE 118-Bus system. The SGOs use this contingency list to defend the attackers and save the SGS from a catastrophe.

The rest of the paper is organized as follows. Section II gives an overview of the Implicative Interdependency Model (IIM) [2] and the Modified Implicative Interdependency Model [1]. Section III describes the K-Contingency list problem and also provides the Integer Linear Programming (ILP) based and heuristic solution for the problem. The simulated solution is also given in section III and the three types of solutions are explained using case studies in this section. Section IV describes the Leader-Follower game theoretic approach followed by adaptive entity hardening to defend different types of attacks on the smart grid system. Performance analysis of the three types of contingency list generation approach and also the adaptive entity hardening technique is discussed in section V of the paper. Finally, the paper is concluded, and future work prospects are given in section VI.

## III. OVERVIEW OF IIM AND MIIM

In both IIM [2] and MIIM [1], the smart grid system can be viewed as a multilayer network, represented as a set $J(E, F(E))$, where $E$ represents set of all entities in both the layers of the smart grid and $F(E)$ represents the set of IDRs. The entities in power layer (layer 1) are considered as P type entities where $P = \{P_1, P_2, ... P_n\}$ and entities in ICT layer (layer 2) are named as C type entities where $C = \{C_1, C_2, ... C_m\}$. The set $F(E)$ is used in both the models to capture the dependencies among interacting entities in the network. Yet, only structural dependencies are considered to generate the IDRs in IIM and both structural as well as operational aspects of the entities are taken into account while formulating IDRs for MIIM. IIM has a binary nature and the entities in that model can either be operational with a state value of 0 or be non-operational with a state value of 1. The most common feature of reduced operability in critical infrastructures is ignored in IIM. The entities in MIIM can take a value of 0, 1 and 2 indicating no-operation, reduced operation and full operation respectively.

Let $C_i$, an entity of layer 2, be operational if (i) $C_j$ which is another entity of layer 2 and $P_a$ which is an entity of layer 1, are operational, or (ii) $C_k$ which is an entity of layer 2 and $P_b$ which is an entity of layer 1 are operational, and (iii) $C_l$ which is an entity in layer 2 is operational. Then the corresponding IIM IDR for $C_i$ would be: $C_i \leftarrow \left((C_j . P_a) + (C_k . P_b)\right) . C_l$. In this IDR, '.' denotes logical AND operation and '+' denotes logical OR operation. Similarly, the IDR for a P type entity can be expressed.

In MIIM, three Boolean operators are used while formulating the IDRs. The first operator is min-AND, denoted by '∘', which selects the lowest of its input values. The second operator is max-OR, denoted by '●', which selects the highest of its input values. The third operator is new_XOR, which is denoted by '◉'. If all the inputs of new_XOR are same, then the output is also same as the inputs. In all other cases the output is 1. This new_XOR operator actually denotes the level of operation of an entity. The truth table for all the 3 new operators are given in Table I.

TABLE I. TRUTH TABLE FOR MIIM OPERATORS

| Input 1 | Input 2 | min-AND | max-OR | new_XOR |
|---|---|---|---|---|
| 2 | 2 | 2 | 2 | 2 |
| 2 | 1 | 1 | 2 | 1 |
| 2 | 0 | 0 | 2 | 1 |
| 1 | 2 | 1 | 2 | 1 |
| 1 | 1 | 1 | 1 | 1 |
| 1 | 0 | 0 | 1 | 1 |
| 0 | 2 | 0 | 2 | 1 |
| 0 | 1 | 0 | 1 | 1 |
| 0 | 0 | 0 | 0 | 0 |

TABLE II. EVALUATION OF IIM AND MIIM IDRs

| | IIM | MIIM |
|---|---|---|
| STEP 1 | $C_l \rightarrow 0$ | $C_l \rightarrow 0$ |
| STEP 2 | $C_i \leftarrow (((2.2) + (2.2)) . 0)$ | $C_i \leftarrow (((2 \circ 2) ● (2 \circ 2)) ◉ 0)$ |
| STEP 3 | $C_i \leftarrow ((2 + 2) . 0)$ | $C_i \leftarrow ((2●2)◉0)$ |
| STEP 4 | $C_i \leftarrow (2 . 0)$ | $C_i \leftarrow (2◉0)$ |
| STEP 5 | $C_i \leftarrow 0$ | $C_i \leftarrow 1$ |

In order to illustrate MIIM, let us assume that if an entity in condition (i) or (ii) fails, $C_i$ will still work full operability, but if (iii) is not satisfied then $C_i$ will operate at a reduced level; this relation can be expressed using MIIM IDRs as: $C_i \leftarrow \left((C_j \circ P_a) ● (C_k \circ P_b)\right) ◉ C_l$. To differentiate between the two models in terms of smart grid system application, the failure of entity $C_l$ for the above IIM and MIIM IDRs are considered and the outcomes are observed in Table II. It is observed in Table II, that for same kind of dependencies, failure of the entity $C_l$ results in the failure of entity $C_i$ in case of IIM but it only reduces the operation level in case of MIIM.

## IV. K-CONTINGENCY LIST PROBLEM

The operator of a smart grid system relies on the sensor-based data like PMU-data and RTU-data to know about the operational state of each and every entity in the power grid. Therefore, it is equally important for the operators to know about the operational states of the communication entities carrying data from the sensors placed in the substations to the control centers. If the operational level of an entity in the system reduces then immediate actions can be taken by the operator. Hence, at a real time, the entities which are more vulnerable to failure should be identified and proper protection or backup to those entities should be provided. This calls the need for an automated system generating the K-Contingency List for the current smart grid system, so that the maximum damage in the power-communication network can be avoided. When one or more entities fail in the smart grid system, many other entities also fail as a result and this is called cascading failures, and this often might lead to a catastrophe if not arrested in time. This cascade stops when the system reaches a steady state once again. Each time a failure takes place in the smart grid, the set $J(E, F(E))$ is updated. All entities that get a state value 0 are removed from the set $E$. As a result, all the IDRs in set $F(E)$ are also updated, since all the dependencies with those failed entities are removed. Now, in between two steady states of the system, there are a number of unstable states of the smart grid when the cascade propagates. Propagation of this cascade may not take place instantly and therefore measures can be taken to arrest the cascade by identifying the K-Contingency List at that time. Given an integer K, and a smart grid system represented as set $J(E, F(E))$, this problem returns the set of K-most critical entities in the joint network, failure of which can lead to the maximum total number of failed entities in the system at the end of the cascade propagation. It is to be noted that a cascade can only propagate in one direction since an already failed entity cannot be affected again by the cascading failure. Therefore, upper bound of the cascade is $|EG| - 1$; where EG is the total number of edges in the network. A formal definition of the problem using the MIIM [1] model is as follows:

### A. Inputs to the Problem
- (a) A joint network $J(E, F(E))$; where $E = P \cup C \cup CP$
  - $P = B \cup T \cup Batt$ (Buses, Transmission Lines/Transformers, Batteries)
  - $C = SE \cup SRE \cup DRE$ (Substation Entities, SONET-Ring Entities, DWDM-Ring Entities)
  - $CP = L \cup R \cup U$ (Power supply lines, RTUs and PMUs)
- (b) Two positive integers K and S.

### B. Decision version of the Problem
Does there exist a set of K entities in E whose failure at time t would result in a failure of at least S entities in total at the next state of the cascading process?

### C. Optimization version of the Problem
Compute the set of K entities in the joint network $J(E, F(E))$ whose failure at time t would maximize the number of entities failed or in other words minimize the overall system state values in the next state of cascade propagation.

The problem of finding K-Contingency List is NP complete, which is already proved in [2]. Therefore, an ILP based solution for the problem is given in section IV and a faster heuristic solution is given in Section V of this paper. Also, validation of the results should be done by comparing the ILP based and heuristic solution results with the simulation results.

### D. Integer Linear Programming (ILP) based solution
In this section, an Integer Linear Programming (ILP) based solution for the K-Contingency List problem stated in Section III of this paper is given. The variable list for the problem is given below—

*1) Variable List:* For each entity $e_i \in E$ a variable set $x_{i,t} \forall t, 0 \le t \le |E| - 1$ is created. The value of $x_{i,t}$ is 2 if it is fully operational, 1 if it is operating at a reduced level of operation and 0 if it is non-operational.

*2) Objective Function:* The objective function for the problem can be defined as:

$$min \sum_{i=1}^{|E|} x_{i,|E|-1} \quad (1)$$

This implies that, the problem aims at minimizing the system states for all the entities in the smart gird.

*3) Constraint Sets:*

*a) Constraint set 1:* $\sum_{i=1}^{|E|} x_{i,0} = K$, entities failed at time step 0 is K.

*b) Constraint set 2:* $x_{i,d} \le x_{i,t-1}$, $\forall t, 1 \le t \le |E| - 1$. This implies that, an entity can only have a system state value at a time $t > d$, less than or equal to the system state value it had at time d.

*c) Constraint set 3:* Based on the 3 new Boolean operations adopted by MIIM, IDRs can have the following format: $e_a \leftarrow (e_b \circledcirc e_c) \circ (e_m \bullet e_n)$.

- Step 1: Firstly, the above IDR can be reformed in the following way: $e_a \leftarrow z_{bcmn}$ where the new variable $z_{bcmn}$ can be expressed as: $z_{bcmn} \leftarrow (g_{bc}) \circ (h_{mn})$ where the two new variables $g_{bc}$ and $h_{mn}$ can be further represented as: $g_{bc} \leftarrow e_b \circledcirc e_c$ and $h_{mn} \leftarrow e_m \bullet e_n$.

- Step 2: Now, a linear constraint is developed for the z type variable (associated with min_AND operator). In order to evaluate the IDR: $z_{bcmn} \leftarrow (g_{bc}) \circ (h_{mn})$, $z_{bcmn}$ can be represented as: $z_{bcmn} \le g_{bc,t-1}$ and $z_{bcmn} \le h_{mn,t-1}, \forall t, 1 \le t \le |E| - 1$.

- Step 3: A linear constraint is also developed for the h type variable (associated with max_OR operator). In order to evaluate the IDR: $h_{mn} \leftarrow e_m \bullet e_n$, $h_{pq}$ can be represented as: $h_{mn} \ge x_{m,t-1}$ and $h_{mn} \ge x_{n,t-1}, \forall t, 1 \le t \le |E| - 1$.

- Step 4: For the g type variable, associated with the new_XOR operator, the following linear constraint is developed. The IDR: $g_{bc} \leftarrow e_b \odot e_c$ is represented by the following set of linear equations: $g_{bc} \geq 0$, $g_{bc} \leq max\_state$, where max_state denotes the state value at the highest level of operability for an entity ( 2 in this case), and $N \times g_{bc} \leq x_{b,t-1} + x_{c,t-1}, \forall t, 1 \leq t \leq |E| - 1$. Here N denotes the number of operands on which the new_XOR operation is taking place.

### E. Heuristic Solution for the problem

The heuristic solution to the self-updating K-Contingency list is completely based on the observations made during the ILP based solutions and simulations.

In order to solve the problem heuristically, first the smart grid system should be considered as a graph $G = (V_P, V_C, E_{PC}, E_{PP}, E_{CC})$ consisting of two different types of vertices $V_P$ and $V_C$ and three different types of edges $E_{PC}, E_{PP}$ and $E_{CC}$. In this abstraction, $V_P$ indicate the power network buses and $V_C$ indicate the communication entities except the channels. All the power or communication channels that connect power and communication entities eg: power supply lines to the communication entities are denoted by $E_{PC}$, Transmission lines and transformers are denoted by $E_{PP}$ and all communication channels are denoted by $E_{CC}$. We are assuming that any edge cannot be most critical as all power networks are (n-1) fault tolerant and all communication networks can adjust routing technique based on failed channels.

1) Initially all the vertices in the graph are considered to be white in color.
2) Input: $G = (V_P, V_C, E_{PC}, E_{PP}, E_{CC})$, K, set of MIIM IDRs and a state table having the state values of each entity.
3) Step 1: The $V_P$ vertices corresponding to generator buses in the actual grid are identified and colored yellow.
4) Step 2: The $V_P$ vertices corresponding to buses with a PMU in the actual grid are identified and colored blue. Any $V_P$ satisfying both the criteria of Step 1 and 2 will be green in color.
5) Step 3: Step 3 solves the problem for K=1
   - Consider a subgraph $G_1 = (V_P, E_{PP})$ ; since a failure of any communication entity cannot bring maximum damage to the smart grid.
   - If the graph has pendant vertices:
     o Identify the pendant $V_P$ vertices.
     o Identify the $V_P$ vertices connected to those pendant vertices and color them Pink.
   - Else if the graph does not have pendant vertices:
     o Identify the $V_P$ vertices having minimum connections.
     o Color those nodes pink.
   - Check the total damage caused by failure of each such pink node by solving MIIM IDRs for those entities only.
   - Select the nodes resulting in maximum damage and color them red.
   - A list of all such red nodes comprise the K=1 contingency list.
   - Change all pink nodes to their previous color.
   - If K=1 then, Go to step 6 else go to step 4.
6) Step 4: Step 4 solves the problem for K=2
   - Take two empty lists List1 and List2.
   - Consider a subgraph $G_1 = (V_P, E_{PP})$ ; since a failure of just two communication entities cannot bring maximum damage to the smart grid.
   - Combine each of the red nodes to each of blue, green and yellow nodes to form all pairs of {Red, Green}, {Red,Yellow} and {Red, Blue}.
   - Check the total damage caused by failure of each such pair by solving MIIM IDRs for those entities in each pair only.
   - Find the {Red, G/Y/B} pair(s) failure of which causes the maximum damage.
     o Add the pair(s) in List1
   - Find all $V_P$ vertices having two $E_{PP}$ edges only.
   - Identify the $V_P$ vertices connected to such $V_P$ vertices having two $E_{PP}$ edges only.
   - Color all such $V_P$ vertices grey.
   - For all such pair of grey $V_P$ vertices:
     o Check the total damage caused by the pair
     o Find the pair(s) causing maximum damage.
     o Add the pair(s) in List2
   - Compare the total damage caused by List1 pairs and List2 pairs.
   - Change all grey nodes back to their previously assigned color.
   - All the pairs causing maximum damage, comprise of the K=2 contingency list.
   - If K=2 then go to step 6, else go to step 5.
7) Step 5: If K>2 this step is executed
   - Round =0, TList1=Empty, TList2=Empty (Round is a counter and TList1 and Tlist2 are two temporary lists)
   - KCon_List =Empty (KCon_List is the K-Contingency List)
   - Graph G2←G1
   - While (TList2 is Empty)

- Find the list of K=2 most vulnerable entities in a list named List_Round (Using step 4 and the input graph G2)
- Remove all the entities in List_Round from the graph G2 and all the connections associated with them.
- Add the pairs in TList1
- If the number of pairs in TList1>=K/2
  - Find all combinations of the pairs in TList1 resulting in a K set.
  - Check the K set causing maximum damage using MIIM IDRs.
  - TList2 ← all such K sets.

- If K is Even
  - KCon_List ← TList2
- Else
  - Consider graph (G1-{Entities in TList2})
  - Convert all the previous red nodes to their last assigned colors.
  - Repeat step 3.
  - Combine TList2 with each current red node obtained.
  - Check the damage caused by solving MIIM IDRs.
  - Find all combinations of TList2 and Red node causing maximum damage.
  - KCon_List ← Each such combinations.

6) *Step 6:* Check if any new failure takes place in the system.
- If yes
  - Update the state values in state table.
  - Remove IDRs of those entities.
  - Remove the entities from the input graph.
  - Repeat step 3 to 6.
- If No
  - Check if there are $V_C$ vertices having all edges $E_{PC}$ connecting them to the $V_P$ entities in the failed list.
  - If yes:
    - Color such $V_C$ vertices red
    - Add such $V_C$ vertices in the K=1 contingency list.
  - Check if there are $V_C$ vertices having all edges $E_{CC}$ connecting them to the $V_C$ entities in the failed list.
  - If yes:
    - Color such $V_C$ vertices red
    - Add such $V_C$ vertices in the K=1 contingency list.
  - The $V_C$ vertices having all edges $E_{PC}$ connecting them to the $V_P$ entities in the contingency list, are also colored red and added to the K=1 contingency list.
  - The $V_C$ vertices having all edges $E_{CC}$ connecting them to the $V_C$ entities in the contingency list, are also colored red and added to the K=1 contingency list.

The main goal of the heuristic solution of the self-updating K-Contingency list is to reduce the search space in order to reduce the computation time of the problem.

*F. Simulated Solution for the problem*

In order to validate the results obtained from the ILP and Heuristic Solutions of the K-Contingency list problem, the smart grid of IEEE 14-Bus system is considered and simulated with various contingencies. To simulate the SGS of IEEE 14-Bus MATPOWER is used for the power network layer and Java Network Simulator is used for the communication network layer. Co-simulation of the two layers is done by passing operation status values of the entities from one layer to the other layer. The same co-simulation platform can be used to simulate larger networks also but finding the K-Contingency list for larger networks by means of simulation and separation of failed entities, is difficult as the problem is NP complete.

*G. Case Studies*

In order to explain the working of the three types of solutions to find the self-updating K-contingency list, a smart grid system of IEEE 14-Bus is considered. In Fig.1, the P type or power entities and the Type 1 communication entities are shown.

Fig. 2. shows the Type 2 communication entities and Fig. 3. shows the Type 3 communication entities. In the smart grid of IEEE 14-Bus system, there are 14 buses and 34 communication terminals like servers, gateways, SADMs and OADMs. It is considered that the transmission lines and communication channels can fail when the entities at the two ends of it also fail. Therefore, IDRs of those entities are not considered. They can either have a state value 1 indicating they are operational or 0 denoting they have failed. However, the other 48 entities (14 *P type* and 34 *C type*) may depend on these transmission lines or communication channels and thus they are included in the IDRs of those 48 entities. Therefore, while finding the K-most vulnerable entities, only 48 entities are taken into account, but those 48 entities also cover the other entities which belong to categories like transmission lines or communication links.

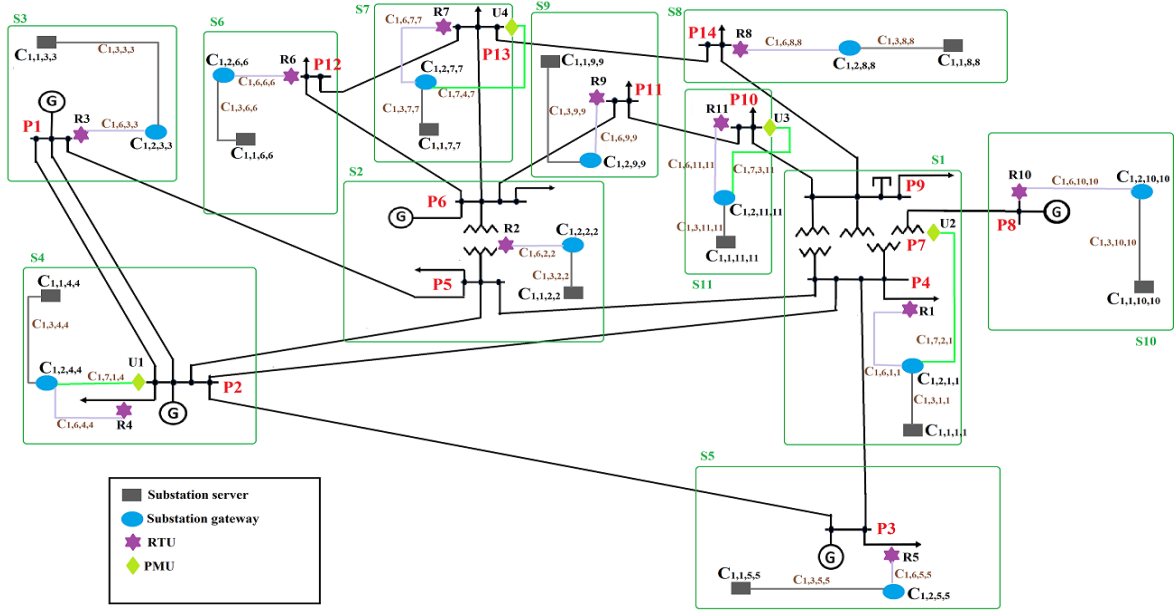

Fig. 1. Power Entities (P) and Type 1 communication entities of a smart grid of IEEE 14-Bus system

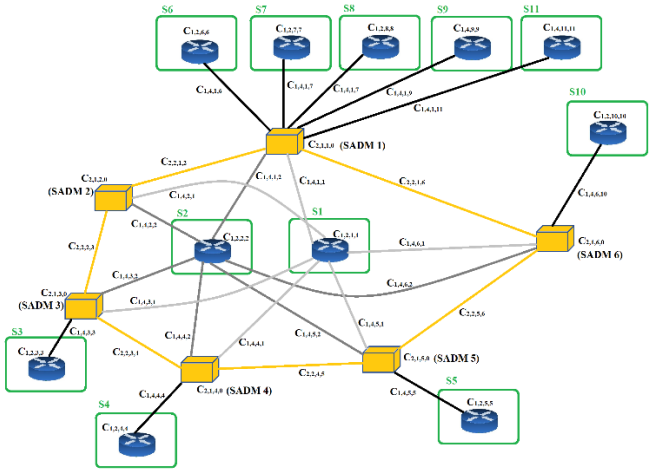

Fig. 2. Type 2 communication entities or SONET-Ring Entities (SRE)

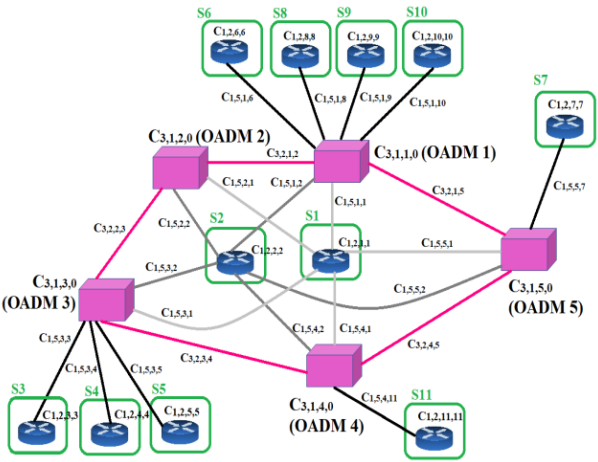

Fig. 3. Type 3 communication entities or DWDM-Ring Entities (DRE)

*1) Case:* Entity $P_{12}$ fails initially

After bus $P_{12}$ located in substation 6 of the smart grid of IEEE 14-Bus fails initially, the contingency list of the system for the next few seconds is analyzed using the MIIM based ILP, IIM based, heuristic solutions and simulation.

    *a) Case a*: ILP based solution using IIM IDRs and MIIM IDRs

TABLE III. SELF-UPDATING CONTINGENCY LIST

| T (ms) | MIIM Contingency List | IIM Contingency List |
|---|---|---|
| 0 | $P_{12}$ fails | $P_{12}$ fails |
| 1 | $\{P_7\}, \{C_{1,2,6,6}\}, \{C_{1,1,6,6}\}$ | $\{P_7\}, \{C_{1,2,6,6}\}, \{C_{1,1,6,6}\}$ |
| 2 | $\{P_7\}, \{C_{1,2,6,6}\}, \{C_{1,1,6,6}\}$ | $\{P_7\}, \{C_{1,2,6,6}\}, \{C_{1,1,6,6}\}, \{C_{2,1,1,0}\}$ |
| 3 | $\{P_7\}, \{C_{1,2,6,6}\}, \{C_{1,1,6,6}\}$ | $\{P_7\}, \{C_{1,2,6,6}\}, \{C_{1,1,6,6}\}, \{C_{2,1,1,0}\}, \{C_{1,2,7,7}\}, \{C_{1,2,8,8}\}, \{C_{1,2,9,9}\}, \{C_{1,2,11,11}\}, \{C_{1,1,7,7}\}, \{C_{1,1,8,8}\}, \{C_{1,1,9,9}\}, \{C_{1,1,11,11}\}$ |
| 4 | $\{P_7\}, \{C_{1,2,6,6}\}, \{C_{1,1,6,6}\}$ | $\{P_7\}, \{C_{1,2,6,6}\}, \{C_{1,1,6,6}\}, \{C_{2,1,1,0}\}, \{C_{1,2,7,7}\}, \{C_{1,2,8,8}\}, \{C_{1,2,9,9}\}, \{C_{1,2,11,11}\}, \{C_{1,1,7,7}\}, \{C_{1,1,8,8}\}, \{C_{1,1,9,9}\}, \{C_{1,1,11,11}\}, \{C_{3,1,1,0}\}, \{C_{3,1,4,0}\}, \{C_{3,1,5,0}\}$ |
| 5 | $\{P_7\}, \{C_{1,2,6,6}\}, \{C_{1,1,6,6}\}$ | $\{P_7\}, \{C_{1,2,6,6}\}, \{C_{1,1,6,6}\}, \{C_{2,1,1,0}\}, \{C_{1,2,7,7}\}, \{C_{1,2,8,8}\}, \{C_{1,2,9,9}\}, \{C_{1,2,11,11}\}, \{C_{1,1,7,7}\}, \{C_{1,1,8,8}\}, \{C_{1,1,9,9}\}, \{C_{1,1,11,11}\}, \{C_{3,1,1,0}\}, \{C_{3,1,4,0}\}, \{C_{3,1,5,0}\}, \{C_{1,2,10,10}\}, \{C_{1,1,10,10}\}$ |

Table III shows the contingency list for 5 ms after $P_{12}$ fails initially, given no new failures take place in the system within this time frame. From the given list below, the K-most vulnerable entities can be selected depending on the K-value. It is also observed that IIM overestimates the number of contingent entities in the network after the initial failure of $P_{12}$.

For K=1, the most vulnerable entity will be $P_7$, for K=2 two sets will be obtained with same priority and any one of the sets can be chosen as K=2 most vulnerable entities. The two sets obtained for K=2 are: $\{P_7, C_{1,2,6,6}\}$ and $\{P_7, C_{1,1,6,6}\}$. This can continue till any K value which is less than the total number of entities in the system at the present state.

*b) Case b:* Heuristic solution using MIIM IDRs

In the heuristic self-updating contingency list solution, after the failure of $P_{12}$, the node corresponding to $P_{12}$ in the input graph is removed and the state table is updated with the current 0 operational value of $P_{12}$. Now, the heuristic algorithm runs from step 3. $P_8$ is the pendant vertex in the graph $G_1 = (V_P, E_{PP})$. Therefore, $P_7$ is the most vulnerable entity in the network.

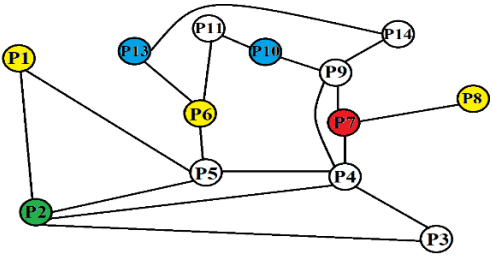

Fig. 4.  K=1 most vulnerable entity for initial failure of $P_{12}$

Now the algorithm executes step 6 and the following entities are added in the list of K=1 contingency list: $\{(P_7), (C_{1,2,6,6}), (C_{1,1,6,6})\}$. Now all these entities are equally vulnerable but for K=1, a P type entity always gets more priority therefore $P_7$ will remain the most vulnerable entity in the system. In Fig.4., the K=1 most vulnerable entity is shown for initial failure of $P_{12}$. Yet, for K=2, if no new failures take place in the system, then the K=2 contingency list will have two pairs of entities–: $\{P_7, C_{1,2,6,6}\}$ and $\{P_7, C_{1,1,6,6}\}$.

*c) Case c:* Simulated Solution

The failure of bus $P_{12}$ is simulated using MATPOWER and Java Network Simulator (JNS). It is observed that after the removal of bus $P_{12}$ from the IEEE 14-Bus system, the power flows converge, and the rest of the network still operates. However, if bus $P_7$ fails then $P_8$ is isolated from the rest of the network and the power flows do not converge in this case. Therefore, the simulated solution also suggests that $P_7$ is the K=1 most vulnerable entity in the SGS. Using JNS, the same C type entities are found in the contingency list as the MIIM based ILP and heuristic solutions. The simulated result also proves that the results obtained using MIIM based ILP and heuristic solutions are valid.

## V. Leader-Follower Technique to Arrest Cascading Failure

In a Smart Grid System (SGS) there can be three types of attackers, namely– intelligent attackers, predictable attackers and unpredictable attackers. Intelligent attackers are humans, predictable attackers can be the nature and unpredictable attackers are the smart grid entities themselves. Intelligent attackers can launch two types of attacks: physical attack and cyber-attack. Predictable attacker can launch only one type of attack which includes all different types of natural disasters and unpredictable attackers can self-destruct any SGS entity.

The leader-follower game also known as Stackelberg game is a game played sequentially between two players [9]. The first player is the leader who commits to the strategy first and then the second player or the follower, commits to his own strategy depending on the strategy of the leader. In such a game, a defender must perpetually defend a set of targets T using a limited number of resources. On the other hand, the attacker may or may not be able to observe and learn the defender's strategy and may attack after careful planning or randomly select targets to damage. In a smart grid scenario, the role of leader or follower can be decided on the basis of the type of attack taking place.

### A. Different Types of Attacks

*1) Game type 1:* Attacks launched by intelligent attackers

*Leader of Game type 1*: In case of attacks launched by intelligent attackers, the Smart Grid Operator (SGO) becomes the leader of this game and also the defender. He selects the most vulnerable entities or the most critical entities in the smart grid beforehand and protect them from attacks or harden them from attacks. The leader also needs to predict the intensions of the attacker and find which of the entities in the network would be an easy target for the attacker. However, the leader has budget constraints and he can only protect a set of entities among all the entities in the smart grid. The mode of protection can be providing a backup device, providing strong security measures for particular entities etc. The type of hardening needed for a particular type of attack is decided by the SGO.

*Follower of Game type 1:* The follower of this game is the attacker. He learns about the strategy of the operator and plans his attack in such a way that he can maximize the damage in the network. Intelligent attackers of the SGS make a careful analysis of the defender's strategy and then come up with a new strategy to maximize the damage in the smart grid system.

*a) Physical Attack:* In case of physical or terrorist attacks, the attacker gets the location details of each substation in the SGS. Then after careful analysis of the defender's strategy to harden the smart grid entities, the intelligent attacks define a new strategy to physically damage particular substations in the SGS in such a way that the overall damage in the system is maximized. These intelligent attackers target those substations where the smart grid entities are not hardened. Example of physical attack can be an Electro Magnetic Pulse (EMP) attack [12].

*b) Cyber Attack:* In cyber-attacks, the target network layer is the communication layer and the entities targeted by the attackers are the ICT entities of the SGS. Just like physical attack, the cyber-attackers are also intelligent attackers and they select the ICT entities in the smart grid in such a way that the overall damage of the communication system can be maximized and as a result, the health monitoring of the critical entities in the SGS can be disrupted. Examples of cyber-attacks can be launching of Denial of Service attack [13], false data injection attack [14] etc.

*2) Game type 2:* Attacks launched by predictable attackers

*Leader of Game type 2*: In game type 2, the attacker acts as the leader. The attacker or mother nature defines her own strategy in which she selects random entities from the SGS and damage them.

*a) Natural Disasters:* The attacker in a type 2 game do not analyze the defender's strategy, rather a particular area in the smart grid is targeted by the attacker and entities in that area are damaged, irrespective of the fact that they are critical entities or not. It is very difficult to defend this attacker as it may or may not have a pattern that can be predicted beforehand. Examples of type 2 attacks causing damage to the smart grid system are– hurricanes, earthquakes etc.

*Follower of Game type 2*: The SGO becomes the follower as well as the defender in this game. The follower in game type 2 tries to predict the strategy of the leader depending on the type of attack the leader wants to launch. For example, if the SGO gets to know that a hurricane is coming, then he first finds the path that will be followed by the hurricane; by the help of weather analysts. Now, based on the path that will be followed by the hurricane, the SGO can predict which of the entities the hurricane can damage and then select the K-most critical entities from that region and harden them before the attacker launches the attack.

*3) Game type 3:* Attacks launched by unpredictable attackers

*Leader of Game type 3*: Just as in game type 2, in game type 3 also the attacker acts as the leader.

Here the attacker can be any entity of the smart grid itself which fails to operate. This attacker does not have a strategy and can attack any entity at any point of time. Therefore, they are unpredictable, and the game starts once the attacker launches an attack on the system. Examples of this type of attacks include damage of any ICT entity or failure of any power entity without any external influence.

*Follower of Game type 2*: The SGO becomes the follower as well as the defender in this game type as well. The follower in game type 3 comes to play once the leader has already started the game by failing one or more entities in the smart grid. Now based on the initial failures, the defender designs his strategy to arrest the cascading failure of entities as well as minimize the damage in the smart grid system.

*B. Description of the Leader-Follower Technique*

In the smart grid system, each target is a smart grid entity. It can either be a P type entity like bus, transmission line, transformer etc. or a C type entity like a server, gateway, communication channel etc. Each target is associated with a set of payoff values that define the utilities for both the defender and the attacker in case of a successful or failed attack [9]. It is assumed in the Stackelberg Security Games, that the payoff of an outcome depends only on the target attacked, and whether that target is hardened by the defender. For example, if an attacker succeeds in attacking a target entity $T_1$ of the smart grid, then the penalty for the defender is same, irrespective of the fact, that some other entity $T_2$ was hardened by the defender or not.

This can be explained using the example in Table IV with only two entities $T_1$ and $T_2$.

TABLE IV. PAYOFF TABLE FOR DEFENDER AND ATTACKER

| Target | Defender | | Attacker | |
|---|---|---|---|---|
| | Hardened | Not Hardened | Hardened | Not Hardened |
| $T_1$ | 2 | 0 | -1 | 1 |
| $T_2$ | 0 | -2 | -1 | 1 |

The payoffs of the security game with only two targets can be shown as in table IV. A set of four payoffs is associated with each target. These four payoffs are the rewards and penalties to the attacker and the defender on the basis of a successful and unsuccessful attack. If a target is attacked, the utility of the defender's utility is given by: $U_D^h(t)$ if the target is hardened, or $U_D^n(t)$ if the target is not hardened. The attacker's utility can also be given in the same way: $U_A^h(t)$ if the target is hardened and $U_A^n(t)$ if the target is not hardened. The table given above shows the utility values. In reality, the $U_D^h(t)$ may correspond to the number of entities that are saved from damage by hardening a target entity and $U_D^n(t)$ may correspond to the number of entities damaged because of a successful attack on a not-hardened entity. Similarly, the utility from the attacker's perspective, $U_A^h(t)$ corresponds to a failed attack with no gain but a penalty of getting detected or penalty of the time devoted or the resources involved to place the attack; and $U_A^n(t)$ may correspond to a successful attack on a not-hardened entity and damage of that entity. It is observed from the table above, that from the defender's perspective, it is always better to harden an entity to gain the maximum utility. On the other hand, from the attacker's perspective, it is always better to have an entity not hardened by the defender so that it can launch the attack and gain a better payoff. However, it may not be feasible to harden all entities in the SGS due to budget constraints.

One approach for the defender would be to find the K-most critical entities in the smart grid system, failure of which can maximize the overall damage of the system at the end of cascading failures. Here K is the budget or the number of entities that the defender can harden at a given time. Now, the defender or the SGO can use K-Contingency list generated using the ILP based or heuristic method to select the smart grid entities which should be hardened. The SGO then hardens those K entities so that the attacker can do no harm to those entities. The hardening approach followed by the SGO is different for different types of entities in the SGS. This is the strategy that the leader of this game will take.

In the similar way, the intelligent attacker can also find the K-most vulnerable entities using the same method as the SGO. It is assumed that the follower or the attacker will know about this strategy and he will know which of the entities in the network are already hardened by the defender. Now, the approach of the attacker should be selecting the M most critical entities in the network where M is the budget of the attacker which denotes the number of entities it can attack at that time. The attacker carefully targets those M entities which should not overlap with the K entities already hardened by the defender. Therefore, if there are E number of entities in the smart grid system, K of them are hardened by the defender, then the remaining entities will be: $(E - K)$. The attacker needs to find the M most critical entities out of the $(E - K)$ entities, the initial failure of which will maximize the damage at the end of the cascading failure process. In this way, the attacker will gain the maximum payoff that he could gain from the current scenario of the network, but the defender will also gain the maximum payoff as he has already hardened the entities, failure of which would have a larger impact on the smart grid.

Now, as the attacker targets the M entities which maximizes the failure of entities in the SGS after K entities are hardened, the next goal of the SGO is to arrest the cascading failure once initial attack on M entities take place.

Similarly, for type 2 attackers which do not have a strategy, the defender tires to predict the M entities out of E entities that might be targeted by the attacker. Then the SGO selects K most critical entities out of those M entities where K<M and hardens those K entities beforehand such that the attacker can possibly harm only $(M - K)$ entities. Thus, protecting the K most vulnerable entities in the targeted region, the defender can arrest the cascading failure of entities in the smart grid.

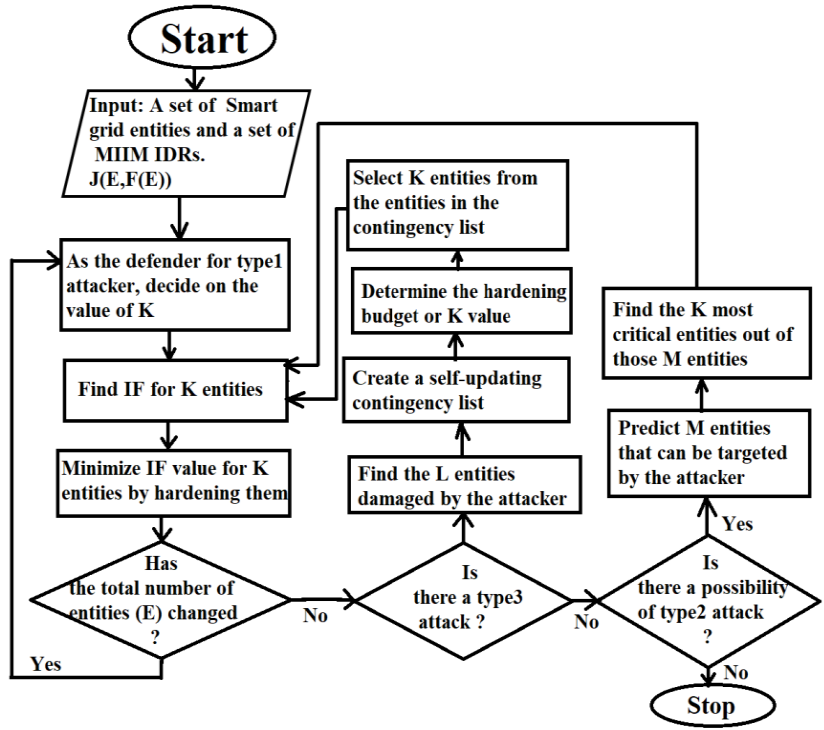

Fig. 5. Flowchart for adaptive hardening of entities

In case of type 3 game, the attacker randomly selects L entities to damage them. They are unpredictable attackers and any preventive measure cannot be taken by the defender. Once the attacker or the leader of this game starts the game by failing L entities, the follower or the defender quickly finds out the list of contingent items that can be harmed as a result. The defender selects K entities out of those $(E - L)$ entities and harden them to arrest the cascading failure of entities in the SGS.

The flowchart given in fig. 5. shows how the cascading failure can be arrested by adaptive hardening of K-most critical entities at a point of time in the SGS. When the K-most vulnerable entities are determined using the MIIM ILP based or MIIM heuristic solution, the impact factor for each such entity in the K-contingency list is determined. Impact Factor (IF) is nothing but the count of the number of entities that will affected as a result of failure of a particular entity. The effect of failure can be change in operational status or complete failure of the affected entity. Now the K-Contingency list is sorted on the basis of this IF value. The entity with the highest IF is selected first. The goal of this method is to arrest the cascading failure by hardening the entity with the highest impact factor. This hardening is done by somehow minimizing the IF value for that entity. Minimizing the IF value for an entity can be done in one or more of the following ways–

- Adjusting generation and load values: if the entity with the highest IF was supplying power to other entities then we can do some load shedding at the buses receiving power from it.

- Adding a backup device for this entity.

- Removing all edges from that entity. That means if the entity with the highest IF is a bus then all transmission channels connecting the entity with the rest of the network is removed and the power flows are adjusted within the rest of the network which now acts as a big separated island. On the other hand, if the entity is a C type terminal entity then all communication paths having that terminal should be avoided for data transmission to the control centers.

## C. Case Studies

With the help of the following case studies, the efficacy of the leader-follower game theoretic approach to arrest cascading failure of smart grid entities can be shown. In order to perform the case studies, a comparatively large smart grid system of IEEE 118-Bus system is considered. The SGS of IEEE 118-Bus is divided into 8 operation zones and 107 substations. Table V shows how the whole SGS is divided into 107 substations and which of the buses are present in which substations. Out of the following 107 substations, substation 61 is selected as the main control center and substation 16 is selected as the backup control center. The basis of control center selection is same as in MIIM [1]. There is a total of 54 SONET Add Drop Multiplexers (SADMs) and 31 Optical Add Drop Multiplexers (OADMs). Fig. 6. shows the different zones in an IEEE 118-Bus system.

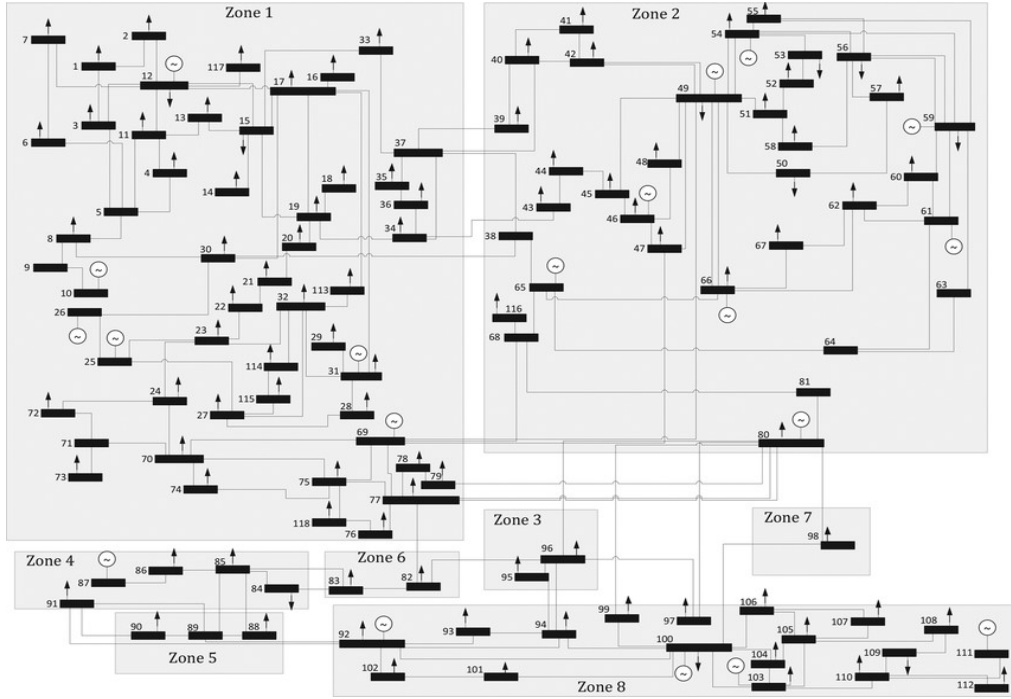

Fig. 6. Zone division of IEEE 118-Bus system

TABLE V.  SUBSTATION DIVISION FOR 118-BUS SMART GRID

| Substation ID | Buses | Substation ID | Buses | Substation ID | Buses | Substation ID | Buses | Substation ID | Buses |
|---|---|---|---|---|---|---|---|---|---|
| 1 | P1 | 23 | P24 | 45 | P49 | 67 | P75 | 89 | P99 |
| 2 | P2 | 24 | P25, P26 | 46 | P50 | 68 | P76 | 90 | P100 |
| 3 | P3 | 25 | P27 | 47 | P51 | 69 | P77 | 91 | P101 |
| 4 | P4 | 26 | P28 | 48 | P52 | 70 | P78 | 92 | P102 |
| 5 | P5, P8 | 27 | P29 | 49 | P53 | 71 | P79 | 93 | P103 |
| 6 | P6 | 28 | P31 | 50 | P54 | 72 | P80, P81 | 94 | P104 |
| 7 | P7 | 29 | P32 | 51 | P55 | 73 | P82 | 95 | P105 |
| 8 | P9 | 30 | P33 | 52 | P56 | 74 | P83 | 96 | P106 |
| 9 | P10 | 31 | P34 | 53 | P57 | 75 | P84 | 97 | P107 |
| 10 | P11 | 32 | P35 | 54 | P58 | 76 | P85 | 98 | P108 |
| 11 | P12 | 33 | P36 | 55 | P59, P63 | 77 | P86, P87 | 99 | P109 |
| 12 | P13 | 34 | P37, P38 | 56 | P60 | 78 | P88 | 100 | P110 |
| 13 | P14 | 35 | P39 | 57 | P61, P64 | 79 | P89 | 101 | P111 |
| 14 | P15 | 36 | P40 | 58 | P62 | 80 | P90 | 102 | P112 |
| 15 | P16 | 37 | P41 | 59 | P65, P66 | 81 | P91 | 103 | P113 |
| 16 | P17, P30 | 38 | P42 | 60 | P67 | 82 | P92 | 104 | P114 |
| 17 | P18 | 39 | P43 | 61 | P68, P69, P116 | 83 | P93 | 105 | P115 |
| 18 | P19 | 40 | P44 | 62 | P70 | 84 | P94 | 106 | P117 |
| 19 | P20 | 41 | P45 | 63 | P71 | 85 | P95 | 107 | P118 |
| 20 | P21 | 42 | P46 | 64 | P72 | 86 | P96 | | |
| 21 | P22 | 43 | P47 | 65 | P73 | 87 | P97 | | |
| 22 | P23 | 44 | P48 | 66 | P74 | 88 | P98 | | |

*1) Case 1: Natural Disaster (Hurricane)*

Case 1 considers a type 2 attack– a hurricane. It is an attack by the type 2 or predictable attacker which do not have any defined strategy. However, the defender or the SGO can know about the path that will be followed by the hurricane from the weather analysts beforehand. Now, the SGO gets to know that a hurricane will pass diagonally through the smart grid region, from the corner of zone 8 towards zone 1. Therefore, the hurricane can affect smart grid entities in zone 8, zone 3 and parts of zone 1. According to the weather analysts, the hurricane can directly damage the following substations–101, 102, 100, 99, 93, 94, 90, 89, 85, 86. Now, the SGO can analyze the effect of the hurricane on the SGS using the MIIM IDRs. By solving the MIIM IDRs, the SGO comes to the conclusion that if the hurricane actually damages the substations mentioned above then many other substations will not be able to send Supervisory Control and Data Acquisition (SCADA) data and Phasor Measurement Unit (PMU) data to the control centers. Table VI shows which of the buses cannot send SCADA and/or PMU data to the control centers due to the damage of the aforementioned substations.

TABLE VI. CONTINGENCY LIST OR LIST OF VULNERABLE ENTITIES PREDICTED BY THE DEFENDER

| Substation ID | Vulnerable Entities |
|---|---|
| 85 | $P_{95}$, |
| 86 | $P_{96}$ |
| 88 | $P_{98}$ |
| 89 | $P_{99}$ |
| 90 | $P_{100}$ |
| 93 | $P_{103}$ |
| 94 | $P_{104}$ |
| 99 | $P_{109}$ |
| 100 | $P_{110}$ |
| 101 | $P_{111}$ |
| 102 | $P_{112}$ |

Now, the defender can find the K-most critical entities out of the list of vulnerable entities given in table VI. If it is assumed that the value of K is 5, then the 5-Contingency list is identified by the defender or SGO in the table VII. The IF value for each entity in the 5-Contingency list is also given in the table.

TABLE VII. 5-CONTINGENCY LIST FOR CASE 1 AND THEIR IF VALUE

| 5-Contingency List | IF value |
|---|---|
| $P_{100}$ | 11 |
| $P_{110}$ | 7 |
| $P_{96}$ | 4 |
| $P_{104}$ | 4 |
| $P_{111}$ | 4 |

Then, the defender hardens these 5 entities in such a way that when the hurricane actually takes place, much lesser number of entities are damaged in the SGS. Fig.7. gives a comparison of the number of operational entities in the SGS if no hardening was done before the hurricane came versus the number of operational entities after the attack when hardening is done. The Fig. 7 also shows the number of operational entities in the normal condition.

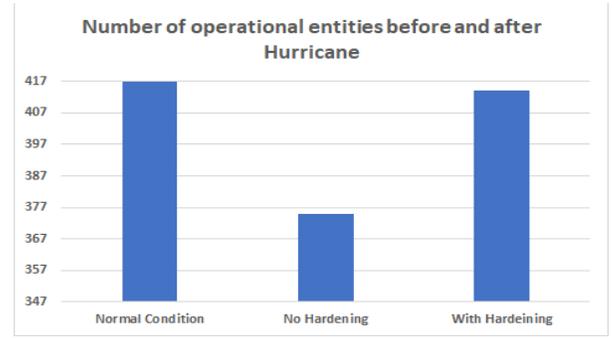

Fig. 7. Number of operational entities before and after the hurricane

*2) Case 2: Electro-Magnetic Pulse Attack*

In this case, a type 1 attack is considered. This type 1 attack is an Electro Magnetic Pulse (EMP) Attack launched by an intelligent attacker with a well-defined strategy. It is to be noted that, in order to arrest the type 2 attack by a predictable attacker, the defender needs to take action after a possibility of attack has raised. The process of defending type 2 attackers is mainly event driven or possibility driven. On the other hand, the SGO or the defender plans on arresting any attack from type 1 attackers from the set-up phase of the SGS. The defender designs his strategy to fail the type 1 attackers right after the SGS is formed and he uses his strategy to defend the attacker whenever a new device is added to the system or when some initial failure has damaged a portion of the SGS. So, in this case, even before any attack is planned by the type 1 attacker, the defender finds the K-most vulnerable entities in the SGS. If it is assumed that the value of K is 5, then the list of following entities are considered in the 5-Contingency list by the SGO.

TABLE VIII. 5-CONTINGENCY LIST FOR CASE 2 AND THEIR IF VALUE

| 5-Contingency List | IF value |
|---|---|
| $P_{68}$ | 301 |
| $P_{69}$ | 301 |
| $P_{17}$ | 90 |
| $C_{1,1,61,61}$ | 299 |
| $C_{1,2,61,61}$ | 299 |

Now, the entities in the contingency list are hardened by the SGO. It is assumed that the intelligent attacker knows about the strategy of the defender and therefore it launches an EMP attack on substation 16 which is the backup control center of the 118-Bus smart grid. In an EMP attack, physical damage of entities take place in the attack location. Therefore, all the entities in substation 16 which includes: buses $P_{17}$ and $P_{30}$; and ICT entities $C_{1,1,16,16}$ and $C_{1,2,16,16}$ should get damaged. Yet, bus $P_{17}$ is hardened by the defender and cannot be damaged by the attacker. Therefore, physical damage of the rest of the entities in substation 16 is done. Now the cascade also spreads from the site of EMP attack, so the defender finds a new set of K-contingency list again to harden them and thereby stop the cascading failure. It is assumed here that for arresting the cascade, the value of K considered is 3. Fig. 8. shows the number of operational entities before the EMP attack, number of operational entities after the EMP attack if no hardening was done and number of operational entities after the EMP attack when adaptive hardening of entities is done.

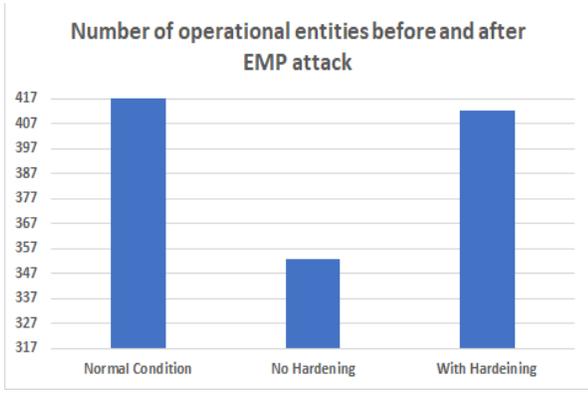

Fig. 8. Number of operational entities before and after the EMP attack

*3) Case 3: Failure of an ICT entity without any external influence*

In this case a type 3 attack is considered. The attacker in this case is gateway of substation 85 ($C_{1,2,85,85}$) which starts the game by self-destroying. The defender of the game gets alert and immediately finds the list of contingent items which may get affected as a result of the failure of ($C_{1,2,85,85}$). It is assumed that the value of K determined by the defender is 3 and table IX shows the list of entities in the 3-Contingency list which are selected for hardening.

TABLE IX. 3-CONTINGENCY LIST FOR CASE 3 AND THEIR IF VALUE

| 3-Contingency List | IF value |
| --- | --- |
| $P_{95}$ | 2 |
| $C_{1,1,85,85}$ | 2 |
| $C_{2,1,46,0}$ | 4 |

Now these entities are hardened by the SGO and fig. 9 shows the number of operational entities before the type 3 attack and after the cascading failure has stopped with hardened and unhardened entities.

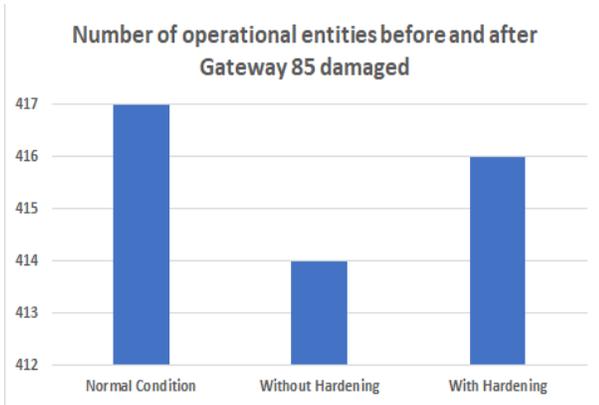

Fig. 9. Number of operational entities before and after Gateway 85 damaged

## VI. PERFORMANCE ANALYSIS

In this section, a comparative analysis of the MIIM [1] ILP and heuristic solution, IIM [2] based K-contingency list solution and simulated solution is done. In order to do the performance analysis of the MIIM model based K-Contingency list identification methods (both ILP and heuristic) and thereby comparing the solutions with that based on IIM model and simulation results, both a small SGS of IEEE 14-Bus and a comparatively large SGS of IEEE 118-Bus is considered in this paper. A co-simulation platform using MATPOWER and Java Network Simulator is used in this paper to simulate the smart grid networks and find the K-Contingency list by simulation method. Java and CPLEX is used to run the MIIM and IIM ILP based approach and only Java is used for the MIIM based heuristic approach to find the self-updating K-Contingency list and also the attack based hardening approach. It is also shown in this section how the performance of the SGS is improved after leader-follower based hardening approach for the smart grid entities is followed.

*A. Number of entities in the contingency list Vs. Time (for initial failure of $P_{12}$) in IEEE 14-Bus SGS*

A type 3 attack is considered here and after bus $P_{12}$ located in substation 6 of the smart grid of IEEE 14-Bus fails initially, the contingency list of the system for the next few seconds is analyzed using the MIIM based ILP and heuristic solutions, IIM based ILP solution and the co-simulation method in fig.10. It is observed that, the simulation results also give the same contingency list as MIIM. It is assumed that no new failures take place even after 5 ms of the failure of bus $P_{12}$. Based on the value of K, the most vulnerable entities in the contingency list are selected.

Now for type 3 game, the defender can only select entities from the contingency list and harden them based on the given budget K and the Impact Factor (IF) value of those entities.

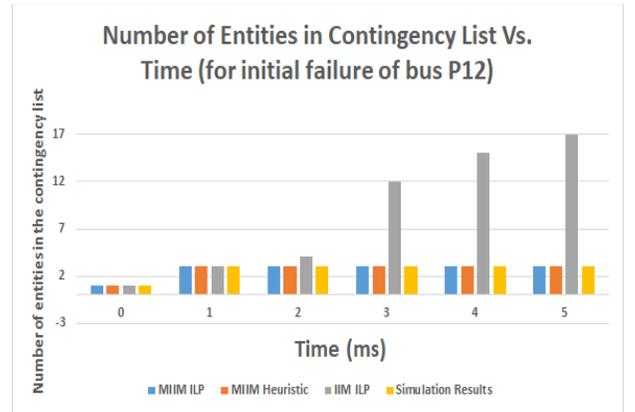

Fig. 10. Number of entities in the contingency list Vs. Time (for initial failure of $P_{12}$)

*B. Number of entities in the contingency list Vs. Time (for initial failure of $P_1$ and $P_{12}$) in IEEE 14-Bus SGS*

Another type 3 attack is considered and in Fig.11 shows the contingency list for MIIM ILP, MIIM Heuristic, IIM ILP and Simulated result after $P_1$. and $P_{12}$. fails initially and no new failures take place even after 8 milliseconds. It is observed that the simulated result of contingency list is same as that obtained using MIIM ILP and MIIM Heuristic. The results obtained using IIM ILP differ a lot from the simulated contingency list. This validates the MIIM model and the ILP and heuristic solution based on MIIM.

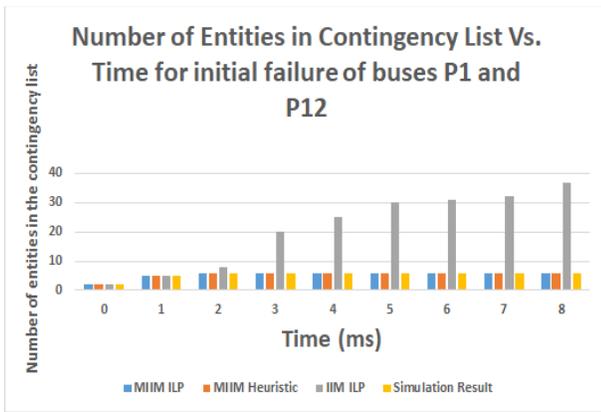

Fig. 11. Number of entities in the contingency list Vs. Time (for initial failure of $P_1$ and $P_{12}$)

### C. Number of entities in the contingency list Vs. Time (for a Type 2 attack) in IEEE 118-Bus SGS

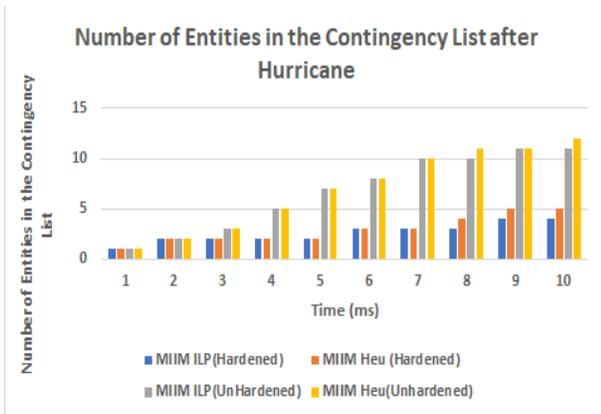

Fig. 12. Number of entities in the contingency list Vs. Time after a Hurricane

A type 2 attack is considered here. A hurricane is predicted to pass over zones 8, 5 and 4; and the SGO gets to know about that. He protects K entities in the contingency list generated on the basis of the prediction. K in this case is 4. Now, as the hurricane actually passes over the three zones, a new self-updating contingency list is generated in real-time to understand which of the entities can get damaged as a result of the attack. Fig. 12 shows a comparison of the MIIM based ILP and heuristic contingency lists for up to 10ms after the hurricane has passed, considering both the situations: (1) region having 4 hardened entities and (2) region having no hardened entities.

### D. Number of entities in the contingency list Vs. Time (for an EMP attack on Substation 45) in IEEE 118-Bus SGS

A type 1 attack is considered in this case. It is assumed that the SGO has already hardened 10 entities out of the 417 entities in the 118-Bus smart grid network. Now, the attacker launches an EMP attack on substation 45, knowing that no entities in that substation is hardened beforehand. Fig.13 shows the number of entities in the MIIM ILP and heuristic solution based contingency list for up to 10ms after the attack took place, considering hardened and unhardened entities in the SGS.

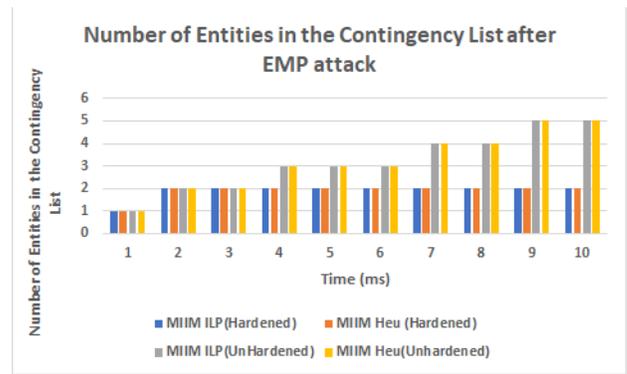

Fig. 13. Number of entities in the contingency list Vs. Time after a Hurricane

It is observed that even after 10 ms, the number of entities in the MIIM ILP as well as heuristic solution based contingency list is the same and consists of 2 vulnerable entities only. On the other hand, there are 5 vulnerable entities in the contingency list when no entities are hardened from beforehand.

It is to be noted that a high K value is given as input for identifying the vulnerable entities after an attack took place.

### E. Maximum Entities damaged vs. K value for both IEEE-14 Bus and IEEE-118 Bus SGS

In fig.14., the maximum damage to the unhardened smart grid network of IEEE 14-Bus after the initial failure of K-most vulnerable entities are predicted by the ILP based solution to the problem using MIIM IDRs and IIM IDRs. Result obtained by solving the problem heuristically using MIIM IDRs is also shown in fig.14. The predicted damages are compared with the simulated results for a smart grid system of IEEE-14Bus.

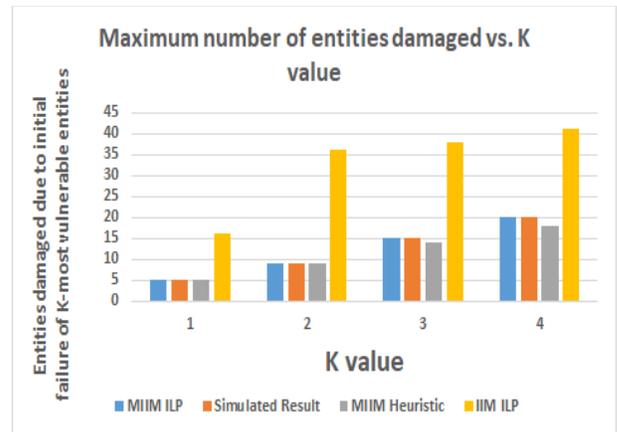

Fig. 14. Maximum number of entities damaged due to the initial failure of K-most vulnerable entities vs. K value (IEEE 14-Bus)

Similarly, in fig.15, the maximum damage to the unhardened and hardened SGS of IEEE 118-Bus after initial failure of the K-most vulnerable entities are shown using MIIM based ILP and heuristic solutions. The hardening is done in the similar manner as type 3 attacks. It is assumed that the K-most vulnerable entities are failing, and the network is not hardened from beforehand. Yet, after the game starts, the defender arrests the failure by hardening that K number of entities and stops the cascade.

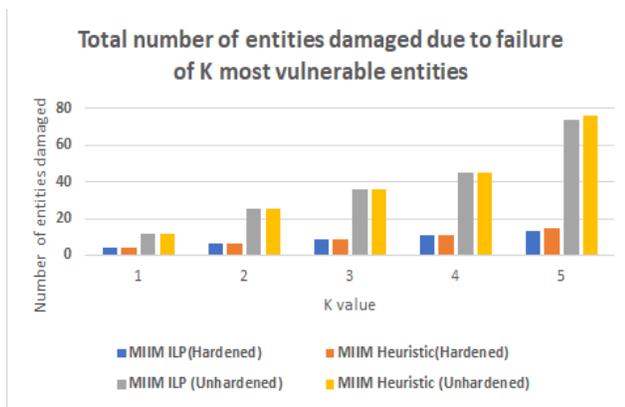

Fig. 15. Maximum number of entities damaged due to the initial failure of K-most vulnerable entities vs. K value (IEEE 118-Bus)

VII. CONCLUSION AND FUTURE WORKS

The Modified Implicative Interdependency Model (MIIM) used in this paper to determine the K-most critical entities works much better than existing interdependency models for critical infrastructure systems. It is observed from the performance analysis that even for larger networks like a smart grid system of IEEE 118-Bus can be protected from several types of attacks using this model and the proposed leader-follower game theoretic approach. In most of the entity hardening based research works, the scientists try to identify the critical entities beforehand and harden them based on their budget. Yet, this approach is not always helpful. Also, contingencies considered and simulated by the smart grid operators before an actual attack may not always match a real scenario. The proposed work considers all the different situations where the defender needs to decide whether he wants to be the leader of the game and take actions beforehand or to play the role of the follower and take actions after an attack is actually launched or a possibility of attack has arrived. The smart grid operator has to modify his strategy accordingly, so that he can protect the maximum number of entities in the smart grid from an attack or a failure. Again, it is proved in the performance analysis that this situation based adaptive hardening method actually performs better and can help in an improved operation of the smart grid system.

The techniques used in this paper to find the K-Contingency list can also be used for progressive recovery [15] of entities in a smart grid system after an attack has taken place in the system. This can be considered as a scope of future work.